\documentstyle[prb,aps,epsf,preprint]{revtex}
\begin{document}
\draft
 
\title{Pressure-induced metallization in solid boron} 

\author{Jijun Zhao $^*$ and Jian Ping Lu $^{\dagger}$}

\address{Department of Physics and Astronomy, University of North Carolina at Chapel Hill, NC 27599-3255}

\date{\today}
\maketitle
\begin{abstract}

Different phases of solid boron under high pressure are studied by first principles calculations. The $\alpha$-B$_{12}$ structure is found to be stable up to 270 GPa. Its semiconductor band gap (1.72 eV) decreases continuously to zero around 160 GPa, where the material transforms to a weak metal. The metallicity, as measured by the density of states at the Fermi level, enhances as the pressure is further increased. The pressure-induced metallization can be attributed to the enhanced boron-boron interactions that cause bands overlap. These results are consist with the recently observed metallization and the associated superconductivity of bulk boron under high pressure (M.I.Eremets et al, Science{\bf 293}, 272(2001)).

\end{abstract}


It is well known that many semiconductors and insulators undergo nonmetal-metal transitions upon applying high pressure. \cite{1,2,3,4,5} Most of these metallization are associated with a structural transition from low coordination insulator to a high coordination metallic phase. \cite{1,3,4,5} In some case, the pressure-induced metallization can be attributed to the increased interatomic interactions which lead to the band overlap. \cite{2}  Recently, pressure-induced metallization and superconductivity were found in solid boron above 160 GPa.\cite{6} There is no direct experimental evidence whether the nonmetal-metal transition is caused by the band gap closure or a structural transition into the metallic bct phase. \cite{7} Moreover, the superconducting transition temperature increases from 6 K at 175 GPa to 11.2 K at 250 GPa. This is in contrast to typical $sp$ metals such as Al, where the $T_c$ decreases with pressure.\cite{8,9} In this letter we present results of first principles calculations on solid boron under high pressure for different phases ($\alpha$-B$_{12}$, bct, and fcc). Our results suggest that $\alpha$-B$_{12}$ phase is stable and maintains its rhombohedral structure up to 270 GPa. Around 160 GPa, a nonmetal-metal transition is found in $\alpha$-B$_{12}$ phase. Further increases in pressure leads to an enhanced density of states at the Fermi level. Charge density analysis shows that the pressure-induced metallization is mainly due to the increased interatomic interactions that broaden the band width. We propose that the recently observed metallization \cite{6} can be understood in terms of the band gap closure. 

Under ambient pressure, solid boron exists in various complex crystal structures with icosahedral B$_{12}$ cluster as a common structural component. The B$_{12}$ icosahedrons can be interlinked by strong covalent bonds in different ways to form different polymorphs such as the $\alpha$-rhombohedral B$_{12}$ ($\alpha$-B$_{12}$), $\alpha$-tetragonal B$_{50}$, or the $\beta$-rhombohedral B$_{105}$ ($\beta$-B$_{105}$). \cite{10,11,15} Among these crystal structures $\alpha$-B$_{12}$ has the simplest form with one B$_{12}$ cluster per rhombohedral unit cell. Previous experimental studies on solid boron include the structural analyses \cite{10}, bulk modulus \cite{11}, Raman spectrum \cite{12}, and electron density distribution \cite{13}. On the theoretical side, the structural, electronic and vibrational properties of solid boron (mostly in $\alpha$-B$_{12}$ phase) have been calculated by first principle methods. \cite{7,12,14,15} In particular, a structural transition from the insulating $\alpha$-rhombohedral to metallic body-centered-tetragonal (bct) structures was predicted to occur at the pressure of 210 GPa. \cite{7} But no detailed analysis on nonmetal-metal transition were done.

\begin{figure}
\centerline{
\epsfxsize=2.5in \epsfbox{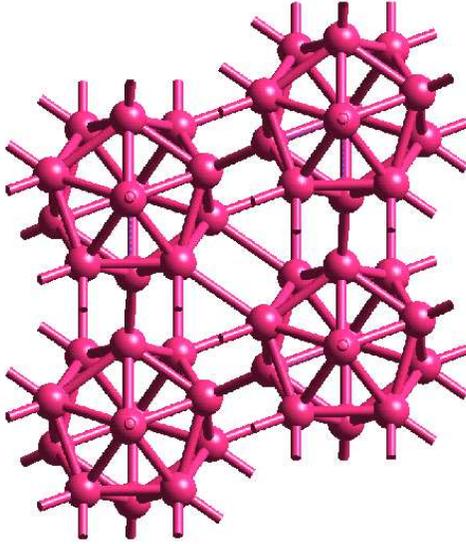}
}
\caption{Optimized $\alpha$-B$_{12}$ rhombohedral crystalline structure shows strong bonding between icosahedral B$_{12}$ clusters. Present GGA calculations give the cell constant $a$=4.98 \AA$~$ and cell angle $\alpha$=58.2$^{\circ}$, which compare well with experimental values of 5.06 \AA$~$, 58.2$^{\circ}$ \cite{10}. (See Table I for the detailed structural parameters.)}
\end{figure}

Our density functional calculations is based on the plane-wave pseudopotential method \cite{16,17}. The density functional is treated by the generalized gradient approximation (GGA) \cite{18} with the exchange-correlation potential parameterized by Wang and Perdew \cite{19}. We use a ultrasoft nonlocal pseudopotential to model the ion-electron interactions \cite{20}. The kinetic energy cutoff for the plane-wave basis is 240 eV. Large set of Monkhorst-Pack {\bf k} meshes \cite{21} ( 8$\times$8$\times$8, 14$\times$14$\times$14, and 14$\times$14$\times$14 for $\alpha$-B$_{12}$, bct, and fcc lattice respectively) are used to sample the Brillouin zone. For a given external pressure, both the cell parameters and the atomic positions are fully optimized.

The optimized crystalline structure of $\alpha$-B$_{12}$ is shown in Fig.1. It is a network of B$_{12}$ clusters with two kinds of intercluster covalent bonds ($d_1$ and $d_2$ in Table I) between neighboring icosahedrons. The calculated cohesive energy 6.95 eV/atom is comparable to the experimental value 5.81 eV. \cite{22} The indirect gap 1.72 eV from our GGA calculation agrees with the estimated experimental value (2.0 eV) \cite{23} better than the previous LDA result (1.43 eV). \cite{14} Table I summarizes the structural parameters of $\alpha$-B$_{12}$ under different pressures. Experimental and previous theoretical results are included for comparison. Satisfactory agreement is found between present calculations and previous works. As shown in the table, upon relaxation the $\alpha$-B$_{12}$ maintains its rhombohedral lattice structure up to high pressure. The cell angle $\alpha$ between the lattice vectors does not change with pressure, while the lattice constant smoothly decreases. All the interatomic bonds are compressed with most notable changes in the long intercluster bonds ($d_2$). This leads to a dramatic modification on the electronic states (to be discussed later).

\begin{table}
Table I. Summary of structural parameters of $\alpha$-B$_{12}$ under different pressures (0 $\sim$ 400 GPa, given in bracket). $a$ denotes the lattice parameter; $\alpha$ describes the angle between the rhombohedral lattice vectors; $r$ is the mean radius of the B$_{12}$ icosahedron; $d_1$ and $d_2$ are the short and long intercluster bond lengths. Available experimental values \cite{10,11} and previous theoretical results \cite{12,15} are included for comparison. 
\begin{center}
\begin{tabular}{cccccc}
                         &  $\alpha$      & $a$ (\AA) & $r$ (\AA) & $d_1$ (\AA) & $d_2$ (\AA) \\ \hline
Exp. \cite{10} (0 GPa)   & 58.2$^{\circ}$ &  5.06     &   1.69    &             &             \\
Exp. \cite{11} (0.5 GPa) & 58.0$^{\circ}$ &  5.07     &   1.70    &  1.67       &  2.00       \\
Theor. \cite{12} (0 GPa) & 58.2$^{\circ}$ &  4.98     &   1.68    &  1.65       &  1.98       \\
Theor. \cite{15} (0 GPa) & 58.1$^{\circ}$ &  5.03     &   1.69    &  1.67       &  2.00       \\
This work (0 GPa)        & 58.2$^{\circ}$ &  4.98     &   1.70    &  1.67       &  1.99       \\
This work (100 GPa)      & 58.0$^{\circ}$ &  4.56     &   1.56    &  1.51       &  1.74       \\
This work (200 GPa)      & 58.1$^{\circ}$ &  4.34     &   1.49    &  1.43       &  1.63       \\
This work (300 GPa)      & 58.2$^{\circ}$ &  4.19     &   1.44    &  1.38       &  1.56       \\
This work (400 GPa)      & 58.3$^{\circ}$ &  4.06     &   1.41    &  1.34       &  1.51       \\
\end{tabular}
\end{center}
\end{table}

\ \\
\ \\
\begin{figure}
\centerline{
\epsfxsize=4.0in \epsfbox{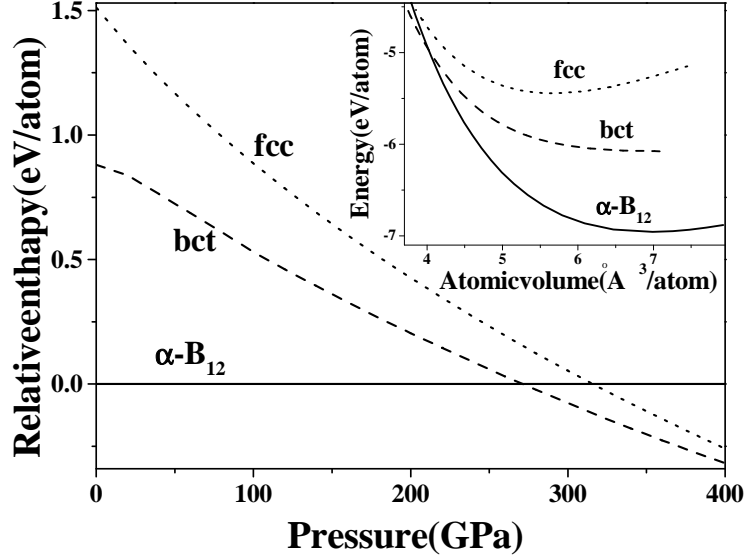}
}
\caption{Relative enthalpy of bct and fcc phase with respect to $\alpha$-B$_{12}$. The crossing of bct and $\alpha$-B$_{12}$ phase occurs at about 270 GPa. No cross between fcc and bct phases is found up to 400 GPa. Insert: cohesive energy per atom of $\alpha$-B$_{12}$, bct, fcc phases as a function of atomic volume.}
\end{figure}

To investigate the pressure-induced structural transition we optimized both the cell parameters and atomic positions for $\alpha$-B$_{12}$, bct, fcc phases under each external hydrostatic pressure. Both the total energy and the enthalpy for each phase were calculated up to 400 GPa. The results are shown in Fig.2. At the ambient pressure, we find 0.82 eV energy difference between the equilibrium $\alpha$-B$_{12}$ and the bct phase, and 1.45 eV difference between the $\alpha$-B$_{12}$ and the fcc phase. Upon applying pressure, $\alpha$-B$_{12}$ remains the low energy structure but the enthalpy of bct phase becomes lower than that of $\alpha$-B$_{12}$ at 270 GPa. Such a crossing indicates a thermodynamic instability of $\alpha$-B$_{12}$ and a possible structural transition to the bct phase. The enthalpy of the fcc phase crosses that of the $\alpha$-B$_{12}$ at the pressure of 315 GPa, but it is never lower than that of the bct phase up to the highest pressure (400 GPa) studied. 

Previous LDA calculations predicted a structural transition from $\alpha$-B$_{12}$ to bct at 210 GPa. \cite{7} We have also preformed independent LDA calculations and obtained a similar transition pressure of 200 GPa. Our GGA transition pressure is about 60 GPa higher than the LDA results. As GGA calculations provide a better agreement with experiments in the band gap than LDA calculations, we believe that the GGA transition pressure (270 GPa) is more reliable. In the previous LDA calculations, a further transition from bct to fcc was predicted to occur at 360 GPa. \cite{7} We have not observed such a transition up to 400 GPa in both LDA and GGA calculations. The discrepancy may due in part to the fact that our calculations are fully optimized in both the lattice parameters and the atomic positions, while the c/a ratio of the bct structure was fixed in the previous calculations. \cite{7}

\begin{figure}
\centerline{
\epsfxsize=3.5in \epsfbox{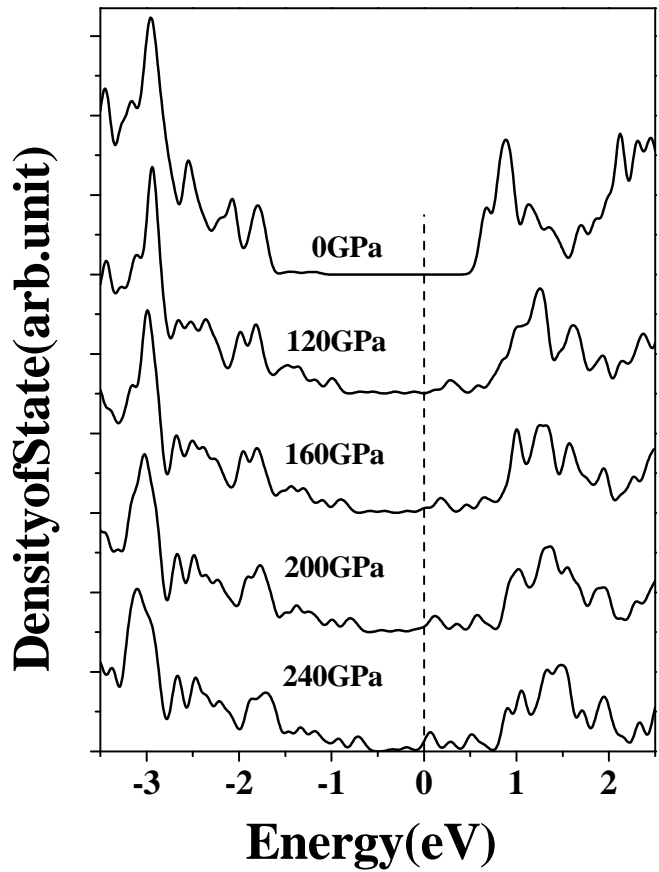}
}
\caption{The electronic density of states of $\alpha$-B$_{12}$ under different pressures. Gaussian broadening of 0.05 eV is used. The broadening of both the valence and the conduction bands with pressure leads to the band gap closure and the formation of metallic states around 160 GPa. }
\end{figure}

From above discussions we conclude that it is unlikely that the experimentally observed nonmetal to metal transition at 160 GPa \cite{6} is due to the structural transition into the metallic bct phase. The alternative explanation for the metallization is the band gap closure upon external pressure, similar to that happened in iodine. \cite{2} As shown in Table I, upon applying pressure, the $\alpha$-B$_{12}$ maintains its rhombohedral lattice structure with compressed intercluster and intracluster bonds. The strengthened boron-boron interactions broaden both the valence and conduction bands, eventually leads to the band gap closure and the development of metallic states near the Fermi level. This effect is clearly shown in Fig.3, where we present the density of states for several different pressures. We perform further analysis of the electron density. Upon applying pressure, both the two-center intercluster bonds and the three-center intracluster bonds \cite{13,14,15} are greatly strengthened while the two-center intracluster bonds are less sensitive. This can be seen in the contour plot of the total charge distributions for different pressures. Fig.4 shows an example of the contour plot, which includes the two-center intercluster and intracluster bonds. A much higher charge density in the area of the intercluster bonds is found at higher pressures. Similar effect was also observed in the three-center intracluster bonds. Thus, we conclude that the enhancement of both intercluster and three-center intracluster bonding leads to pressure-induced metallization. 

\begin{figure}
\centerline{
\epsfxsize=3.0in \epsfbox{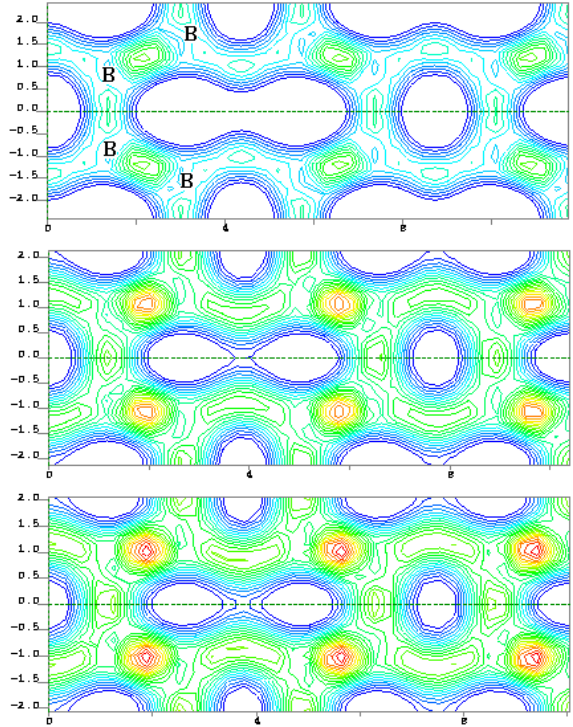}
}
\caption{The total charge density contour plot for $\alpha$-B$_{12}$ phase under different pressures: 0 GPa (upper), 160 GPa (middle), 240 GPa (down). The current slice includes two-center intercluster and intracluster bonds. The representative sites for boron atoms are labeled as B in upper plot. Red, yellow, green, blue colors indicate decreasing electron density. A significantly enhanced charge density (red color) in the area of the intercluster bond can be seen in the case of higher pressures. }
\end{figure}

A clearer picture of nonmetal-metal transition can be seen in the pressure dependent of the band gap shown in Fig.5. From extrapolation, the band gap closure occurs around 160 GPa. Experimentally, it was found that $\beta$-B$_{105}$ undergo a nonmetal to superconductor transition at about 160 GPa \cite{6}. Although our calculations were based on $\alpha$-B$_{12}$ phase, we believe that our results are qualitatively comparable to the experiments because both phases have the same icosahedral structural component. 
\ \\
\ \\
\ \\
\ \\
\begin{figure}
\centerline{
\epsfxsize=3.5in \epsfbox{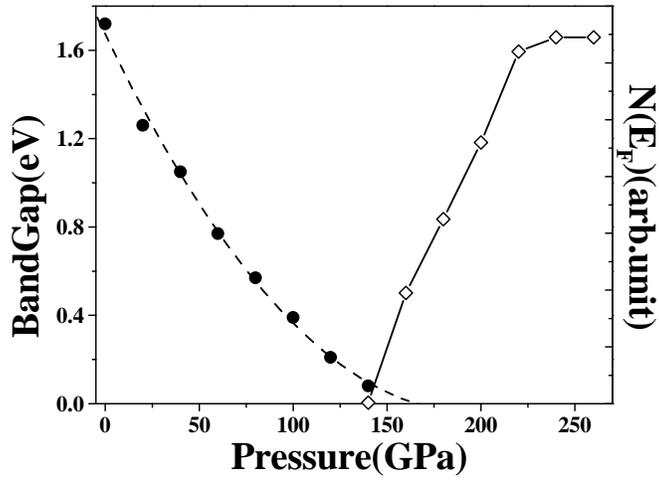}
}
\caption{The band gap (filled dots) and the electronic density of states $N(E_F)$ (open squares) at the Fermi level for $\alpha$-B$_{12}$ as functions of pressure. The extrapolated pressure for band gap closure is 160 GPa. In the metallic phase, $N(E_F)$ increases rapidly with pressure at first, then tampers off around 240 GPa. }
\end{figure}

In the metallic phase of $\alpha$-B$_{12}$ our calculations show that density of states $N(E_F)$ at the Fermi level increases rapidly with pressure from 160 GPa to 240 GPa, beyond which it saturates (see Fig.5). The enhancement of metallicity with pressure can be directly associated with the increase of superconducting transition temperature $T_c$ from 6 K at 175 GPa to 11.2 K at 250 GPa.\cite{6} In the phonon mediated superconductivity, the $T_c$ can be obtained as \cite{24,25}$$T_c=\frac{\omega_{ln}}{1.2}exp(-1.04\frac{1+\lambda}{\lambda -\mu^*-0.62\lambda \mu^*}),$$ where $\omega_{ln}$ is the characteristic phonon frequency, $\lambda$ is the electron-phonon coupling strength, $\mu^*$ is the screened coulomb repulsive interaction. The electron-phonon coupling $\lambda$ is related to the density of states at the Fermi level $N(E_F)$ by 
$$
\lambda=N(E_F)<I^2>/(M<\omega^2>),
$$
where $<I^2>$ is the average square of the electron-phonon matrix element and $<\omega^2>$ is averaged phonon frequency.\cite{25} Typically for $sp$ metal such as aluminum, $<\omega^2>$ increase with pressure and $N(E_F)$ decreases, while $\mu^*$ is insensitive. \cite{9} Thus, $T_c$ decreases with pressure. \cite{8,9}  On the contrary, in solid boron, $T_c$ was found to increases substantially with pressure (175$\sim$250 GPa). \cite{6} Our calculations show that such increase in $T_c$ can be understood by the dramatic increase in the density of states $N(E_F)$ with pressure after the metallization of $\alpha$-B$_{12}$. In addition, we also have calculated the $N(E_F)$ and its pressure dependence for the metallic bct phase. The $N(E_F)$ for bct phase is much higher than that of $\alpha$-B$_{12}$ phase and it does not show a clear pressure dependence. This result further supports our interpretation that the observed metallization in boron may not originate from a structural transition from $\alpha$-B$_{12}$ to bct phase. 

In summary, we have performed first principles calculations to investigate the structural and electronic properties of solid boron under high pressure. Structural transition from $\alpha$-B$_{12}$ to bct phase is found to occur at about 270 GPa. The semiconductor gap (1.72 eV at ambient pressure) decreases upon external pressure. Band gap closure and a nonmetal-metal transition is found around 160 GPa. As the external pressure further increase (160 GPa to 240 GPa), the density of states at Fermi level increase dramatically. We conclude that the experimentally observed metallization in boron might be due to the band gap closure instead of a structural transition into the metallic bct phase. 

This work is supported by the U.S. Army Research Office (Grant DAAG55-98-1-0298). We acknowledge computational supports from the North Carolina Supercomputer Center.
\ \\
$^*$: zhaoj@physics.unc.edu \\
$^{\dagger}$: jpl@physics.unc.edu


\vspace{0.5cm}
\begin{references}

\bibitem{1} J.C.Jamieson, Science{\bf 139}, 129(1963); J.Wittig and B.T.Matthias, Science{\bf 160}, 92(1968).

\bibitem{2} O.Shimomura, K.Takemura, Y.Fujii, S.Minomura, M.Mori, Y.Noda, Y.Yamada, Phys.Rev.B{\bf 18}, 715(1978).

\bibitem{3} K.J.Dunn, F.P.Bundy, J.Appl.Phys.{\bf 51}, 3246(1980)£" F.P.Bundy, K.J.Dunn, Phys.Rev.B{\bf 22}, 3157(1980).

\bibitem{4} A.Jayaraman, Red.Mod.Phys.{\bf 55}, 65(1983).

\bibitem{5} G.J.Ackland, Rep.Prog.Phys.{\bf 64}, 483(2001).

\bibitem{6} M.I.Eremets, V.V.Struzhkin, H.K.Mao, R.J.Hemley, Science{\bf 293}, 272(2001).

\bibitem{7} G.Mailhiot, J.B.Grant, A.K.McMahan, Phys.Rev.B{\bf 42}, 9033(1990).

\bibitem{8} D.U.Gubser, A.W.Webb, Phys.Rev.Lett.{\bf 35}, 104(1975).

\bibitem{9} M.M.Dacorogna, M.L.Cohen, P.K.Lam, Phys.Rev.B{\bf 34}, 4865(1986).

\bibitem{10} G.Will, B.Keifer, B.Morosin, G.A.Slack, Mater.Res.Soc.Symp.Proc.{\bf 97}, 151(1987).

\bibitem{11} R.J.Nelmes, J.S.Loveday, D.R.Allan, J.M.Besson, G.Hamel, P.Grima, S.Hull, Phys.Rev.B{\bf 47}, 7668(1993).

\bibitem{12} N.Vast, S.Baroni, G.Zerah, J.M.Besson, A.Polian, M.Grimsditch, J.C.Chervin, Phys.Rev.Lett.{\bf 78}, 693(1997).

\bibitem{13} M.Fujimori, T.Nakata, T.Nakayama, E.Nishibori, K.Kimura, M.Sakata, Phys.Rev.Lett.{\bf 82}, 4452(1999).

\bibitem{14} S.Lee, D.M.Bylander, L.Kleinman, Phys.Rev.B{\bf 42}, 1316(1990).

\bibitem{15} D.Li, Y.N.Xu, W.Y.Ching, Phys.Rev.B{\bf 45}, 5895(1992).

\bibitem{16} The density functional plane-wave pseudopotential calculations are performed by using CASTEP. CASTEP is an {\em ab initio} program with plane-wave pseudopotential package distributed by Accelrys Inc: http://www.accelrys.com/.

\bibitem{17} M.C.Payne, M.T.Teter, D.C.Allen, T.A.Arias, J.D.Joannopoulos, Rev.Mod.Phys.{\bf 64}, 1045(1992).

\bibitem{18} J.P.Perdew and Y.Wang, Phys.Rev.B{\bf 45}, 13244(1992).

\bibitem{19} Y.Wang and J.P.Perdew, Phys.Rev.B{\bf 43}, 8911(1991).

\bibitem{20} D.Vanderbilt, Phys.Rev.B {\bf 41}, 7892(1990).

\bibitem{21} H.J.Monkhorst, J.D.Pack, Phys.Rev.B{\bf 13}, 5188(1976).

\bibitem{22} C.Kittle, {\em Introduction to Solid State Physics}, 7th ed., (Wiley, New York, 1996).

\bibitem{23} F.H.Horn, J.Appl.Phys.{\bf 30}, 1611(1959).

\bibitem{24} P.B.Allen, R.C.Dynes, Phys.Rev.B{\bf 12}, 905(1975).

\bibitem{25} W.L.McMillan, Phys.Rev.{\bf 167}, 331(1968).

\end{references}
\end{document}